# The Generation of Large Networks from Web-of-Science Data

Loet Leydesdorff,[*a] Gohar Feroz Khan,[b] and Lutz Bornmann[c]

[*] Corresponding author
[a] Amsterdam School of Communication Research (ASCoR), University of Amsterdam, PO Box 15793, 1001 NG Amsterdam, The Netherlands; loet@leydesdorff.net .

[b] Korea University of Technology & Education (KoreaTECH), 1600 Chungjol-ro Byungcheon-myun Cheonan city, 330-708, South Korea; gohar.feroz@kut.ac.kr

[c] Division for Science and Innovation Studies, Administrative Headquarters of the Max Planck Society, Hofgartenstr. 8, 80539 Munich, Germany; bornmann@gv.mpg.de .

**Abstract**

During the 1990s, one of us developed a series of freeware routines (http://www.leydesdorff.net/indicators) that enable the user to organize downloads from the Web-of-Science (Thomson Reuters) into a relational database, and then to export matrices for further analysis in various formats (for example, for co-author analysis). The basic format of the matrices displays each document as a case in a row that can be attributed different variables in the columns. One limitation to this approach was hitherto that relational databases typically have an upper limit for the number of variables, such as 256 or 1024. In this brief communication, we report on a way to circumvent this limitation by using txt2Pajek.exe, available as freeware from http://www.pfeffer.at/txt2pajek/.

**Introduction**

In recent decades, one of us has developed a series of software routines (http://www.leydesdorff.net/indicators) that enables the user to organize downloads from the Web-of-Science (Thomson Reuters) into a relational database, and then to export matrices for further analysis in various formats; for example, for co-author analysis, co-citation analysis, bibliographic coupling, etc. (Cobo *et al.*, 2011). The basic format of each matrix shows each document as a case in a row that can be attributed with different variables in the columns. Variables can be author names, institutional addresses, cited references, etc. One can also combine types of variables such as authors, title words, and institutional addresses (Leydesdorff, 2014; Vlieger & Leydesdorff, 2011). Multiplication of the asymmetrical word/document matrix with its transposed leads to a co-word matrix; and this operation can be done *mutatis mutandis* for other (sets of) variables attributable to documents.

One limitation to this approach was hitherto that relational databases typically have an upper limit for the number of variables, such as 256 or 1024, whereas the number of cases (documents) is limited only by considerations of disk space.[1] In this brief communication, we report on a way to circumvent this limitation easily by using txt2Pajek.exe, available as freeware from http://www.pfeffer.at/txt2pajek/ (Pfeffer *et al.*, 2013). Txt2Pajek enables the user to generate a 2-mode (asymmetrical) matrix of cases (documents) and variables in the Pajek format for an unlimited number of variables from a text file. Within Pajek (de Nooy *et al.*, 2011) the newly generated 2-mode file can be further transformed into a 1-mode network file that can also be used in other software programs for network analysis and visualization such as Gephi, UCINet, or VOSViewer.

---
[1] In a 32-bit operating environment, file sizes are limited to 2 GB, but this limitation is removed in the environment of a 64-bit operating system.

**Data**

One of us (GFK) encountered the systems limitation of 1024 variables when generating a co-author network at the level of institutional addresses using instcoll.exe (at http://www.leydesdorff.net/software/instcoll/index.htm) for analysis and visualization. Using the eight journals listed in the so-called Senior Scholars' Basket of the Association for Information Systems (AIS) that were used for the ranking, 3,587 documents were downloaded for the period 1995-2014 (Table 1). The set contains 7,397 institutional addresses, of which 4,617 are unique (Khan, in preparation). The author wished to pursue a network analysis using these names of institutions as nodes and had already organized the data download in a relational database using isi.exe (at http://www.leydesdorff.net/software/isi).[2]

| Journals | N |
|---|---|
| *European Journal of Information Systems* | 613 |
| *Information Systems Journal* | 341 |
| *Information Systems Research* | 549 |
| *Journal of Information Technology* | 447 |
| *Journal of Management Information Systems* | 534 |
| *Journal of Strategic Information Systems* | 331 |
| *Journal of the Association For Information Systems* | 239 |
| *MIS Quarterly* | 533 |
| **Total** | **3,587** |

**Table 1**: The data of 3,587 documents in the eight journals (1995-2014) in the basket used by the Association for Information Systems (AIS) for ranking.

---

[2] One can use scopus.exe at http://www.leydesdorff.net/scopus/ for transforming data from Scopus into this format.

**Analysis**

The institutional names are organized by isi.exe in a separate table (named "cs.dbf") that contains the document numbers for relational database management and the address information. Using Excel or a similar program, one can open this table and save it as a comma-separated-variables (.csv) file or as tab-delimited. A program entitled dbf2csv.exe has additionally been made available at http://www.leydesdorff.net/software/dbf2csv/dbf2csv.exe for a direct transformation. The comma-separated files can be read as text files into txt2Pajek.exe and are transformed in 2-mode Pajek files.

One can further refine the address information by using functions of Excel. For example, the first address in the file was "UNIST, Sch Technol Management, Ulsan, South Korea" in cell B2. Using the function "=left(B2, find(",",B2)-1)", one obtains the institutional name "UNIST" in another cell (e.g., C2). Since institutional names are now considerably standardized in WoS, one can drag the function along the column in Excel and thus obtain a field with only institutional names. There are 1,364 unique institutional names in the set based on 3,564 (of the 3,578) documents. Similarly, one can extract country names on the right side of the string using more composed functions or by writing a routine.[3]

The .csv file should be renamed with the extension ".txt" and one should take care that the content is either lower or upper case (or capitalized case) because the default cases were changed in WoS during the 1990s. The transformation by txt2Pajek is straightforward and provides a file with the same name, but with the extension ".net" in the Pajek format. This file can be read into Pajek or another network-analysis program that is able to read this format. The Pajek format is nowadays increasingly the standard currency for exchanges among network analysis and visualization programs.

---

[3] The table cs.dbf already contains country names as a separate (third) field.

**Network analysis in Pajek**

When the 2-mode network generated by txt2pajek.exe is read into Pajek (v.3), it can be transformed into a 1-mode network (in this case of institutes) under *Network > 2-Mode Network > 2-Mode to 1-Mode > Columns*. The option "multiple lines" should be set ON. Thereafter the multiple lines have to be summed under *Network > Create New Network > Transform > Remove > Multiple Lines > Sum Values.* The lines of the network (edges) can now be visualized with different widths. Similarly, one can size the nodes using "weighted degrees" for the number of occurrences under *Network > Create Vector > Centrality > Weighted Degree > All.* Using the Draw-menu now visualizes the network (*Draw > Network + First Partition + First Vector*; Bruuns, 2009).

As would be expected, institutional collaboration networks contain lots of isolates, dyads, triads, etc. These small networks are not necessarily connected among themselves. The network under study thus contains 153 components with a largest component of 1,171 (85.9% of the 1,364) nodes. Figure 1 shows this largest component as a heat map after exporting to VOSViewer (Van Eck & Waltman, 2010). The second largest component contains only nine institutes.

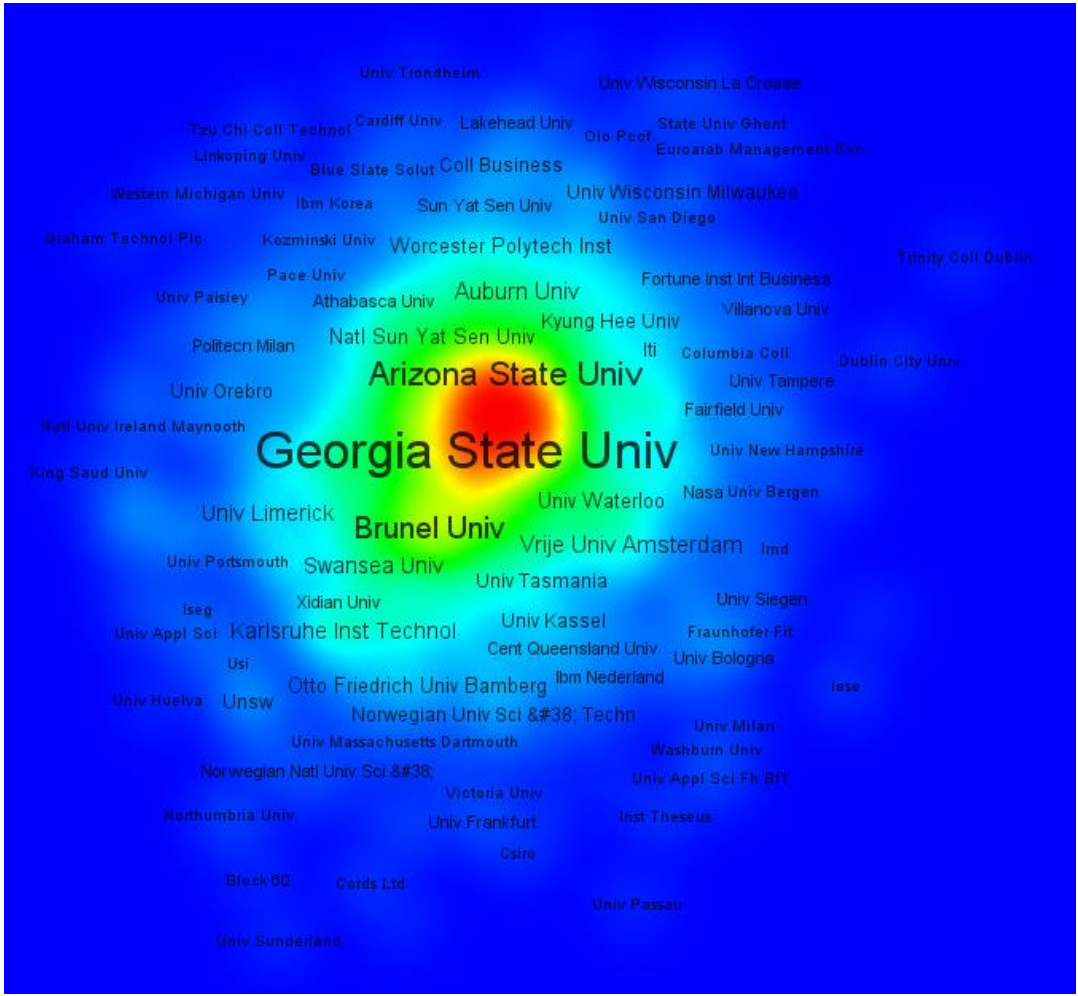

**Figure 1**: Heat map of the largest component ($N = 1,171$) of the network of institutional collaborations in the AIS-basket of 8 journals in "information systems".

Figure 2 shows the network of 67 countries named in the institutional addresses of the 3,563 (99.6%) documents that provide such information. Note that in WoS, "England" is counted separately from the other countries of the UK.

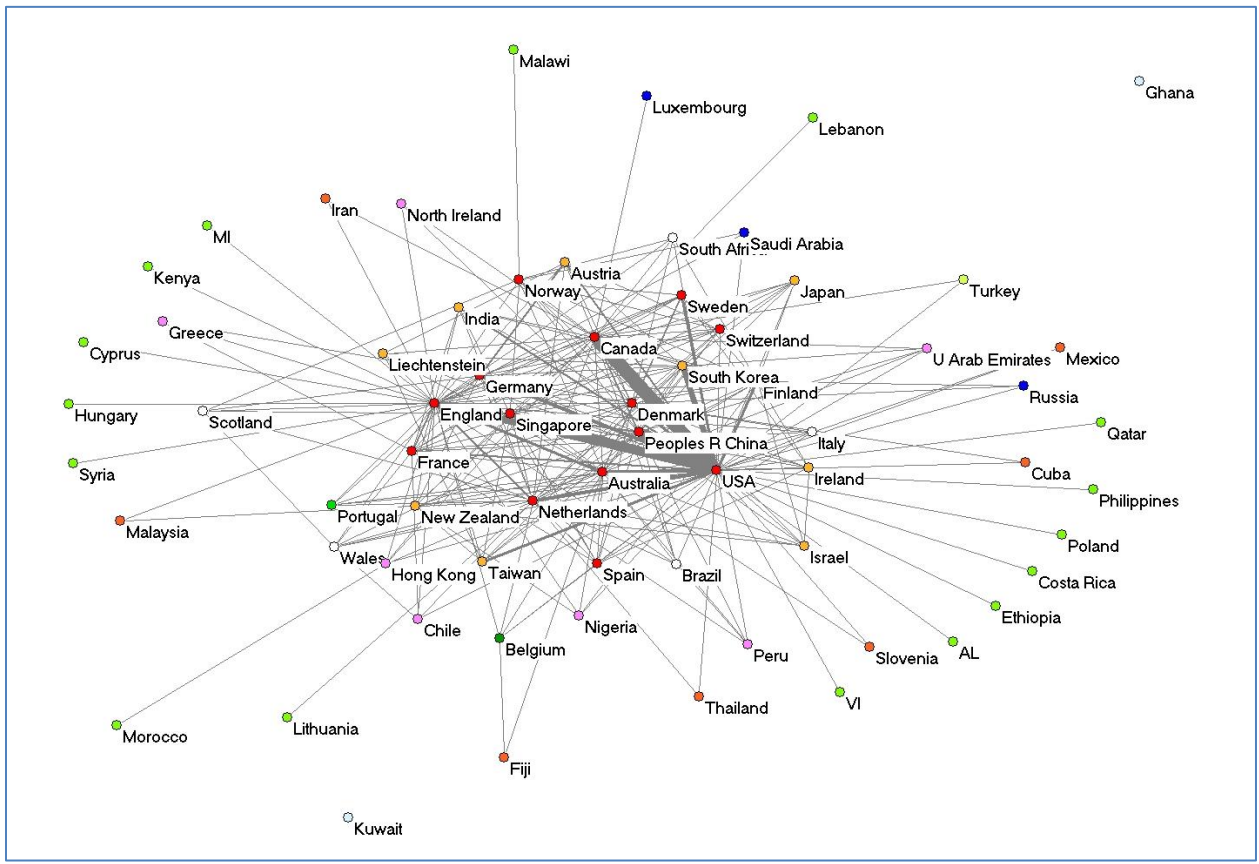

**Figure 2:** Network of international collaborations in the AIS basket; 67 countries; 3,563 documents. Kamada & Kawai (1989) used for the mapping.

**Conclusions and summary**

Using this pathway, one can visualize both smaller and very large networks, for example, of authors in large consortia (such as at CERN; Milojevic, 2010). The routines isi.exe and txt2pajek.exe have no systems limitations except disk sizes. Bringing the files into network analysis and visualization programs, one can study degree distributions, clustering coefficients, modularity, etc., and visualize subsets accordingly. An alternative route for achieving this is provided by Wos2Pajek available at http://pajek.imfm.si/doku.php?id=wos2pajek , but in this case the data is not organized relationally into databases. We have demonstrated the possibilities for analysis and visualization of collaborations in the specialty of "information systems" both at the institutional and international levels.